\documentclass[11pt,dvips]{article}
\usepackage{rotating}
\usepackage{hhline}
\usepackage{array}
\usepackage{epsfig}
\usepackage{amssymb}

\advance\textheight by \topskip
\begin{document}
\tolerance=100000
\thispagestyle{empty}
\setcounter{page}{0}

\renewcommand{\refname}{}

\newcommand{\nn}{\nonumber}
\newcommand{\be}{\begin{equation}}
\newcommand{\ee}{\end{equation}}
\newcommand{\ba}{\begin{eqnarray}}
\newcommand{\ea}{\end{eqnarray}}
\newcommand{\bann}{\begin{eqnarray*}}
\newcommand{\eann}{\end{eqnarray*}}
\newcommand{\bc}{\begin{center}}
\newcommand{\ec}{\end{center}}
\newcommand{\ds}{\displaystyle}
\newcommand{\half}{\frac{1}{2}}
\newcommand{\sts}{\scriptstyle}
\newcommand{\ngs}{\!\!\!\!\!\!}
\newcommand{\rb}[2]{\raisebox{#1}[-#1]{#2}}
\newcommand{\CP}{${\cal CP}$~}
\newcommand{\cosb}{\cos \beta}
\newcommand{\sinb}{\sin \beta}
\newcommand{\ml}{\frac{1}{\sqrt{2}} \lambda v}
\newcommand{\sbomu}{\frac{\sin 2 \beta}{2 \mu}}
\newcommand{\kmol}{\frac{\kappa \mu}{\lambda}}
\newcommand{\mytitle}[1]{\begin{center}
{\bf \underline{#1}} \end{center}}
\newcommand{\msta}{m_{\tilde{t}_1}}
\newcommand{\mstb}{m_{\tilde{t}_2}}
\begin{titlepage}

\begin{flushright}
CERN-TH/2003-209
\end{flushright}

\vspace{1cm}

\renewcommand{\thefootnote}{\fnsymbol{footnote}}
\begin{center}
  {\Large \bf The Peccei-Quinn Axion in \\ 
the Next-to-Minimal Supersymmetric Standard Model}\\[1cm]
{\large D.J.~Miller$^{1}$ and  R.~Nevzorov$^{2}$} \\[1cm]
 {\it 
$^1$ Theory Division, CERN, CH-1211 Geneva 23, Switzerland\\
$^2$ ITEP, Moscow, Russia}\\
\end{center}

\renewcommand{\thefootnote}{\arabic{footnote}}
\vspace{3cm}

\begin{abstract}
\noindent
We discuss the Next-to-Minimal Supersymmetric Standard Model (NMSSM)
with a Peccei-Quinn (PQ) $U(1)$ symmetry. When this symmetry is
dynamically broken by the Higgs mechanism, the resulting
pseudo-Nambu-Goldstone boson takes the role of an axion. Although much
of the allowed parameter space for low values of the PQ scale has been
ruled out, many scenarios with a PQ scale $\gtrsim 10^9 \; {\rm GeV}$
remain untested, allowing the NMSSM PQ axion to provide a solution to
the strong CP problem and be a good dark matter candidate.
Unfortunately the new particle states are so decoupled that they would
not be observable at future colliders, and the NMSSM would appear
indistinguishable from the minimal model.  However, we show that in
order to maintain vacuum stability, such a model requires that the
heavy Higgs boson states have masses that lie close to approximately
$\mu \tan \beta$. Therefore, a measurement of the Heavy Higgs boson
masses at the LHC would allow one to either rule out the NMSSM PQ
axion, or provide tantalizing circumstantial evidence for its
existence.
\end{abstract}

\end{titlepage}

\mytitle{Introduction: The Strong CP Problem and the Axion}

For some some time after its formulation, one of the principle
strengths of Quantum Chromodynamics (QCD) was thought to be its
automatic conservation of parity (P) and charge-conjugation--parity
(CP) symmetries. The only renormalizable P and CP violating term that
may be added to the QCD Lagrange density is the ``$\theta$-term'',
\be {\cal L}_{\theta} = \theta_{\rm eff} \frac{\alpha_s}{8 \pi} 
F^{\mu \nu \, a} \tilde{F}_{\mu \nu}^a, \label{eq:thetaterm} \ee
where $F^{\mu \nu \, a}$ is the gluon field strength and
$\tilde{F}_{\mu \nu}^a \equiv \frac{1}{2} 
\epsilon_{\mu\nu\rho\sigma}F^{\rho\sigma\, a}$ is it dual;
$\theta_{\rm eff}$ is the effective $\theta$-parameter after 
diagonlization of the quark mass matrix, 
i.e.\ $\theta_{\rm eff} = \theta + {\rm arg \; det}\; M_q$.
It is straightforward to show that this term is a total derivative
allowing its integral over space-time to be written as a boundary term
at infinity.  Therefore, it was thought, its integral will vanishes in
the vacuum, and the $\theta$-term may be safely ignored.

However, it was soon realized that such a term could not be ignored if
the vacuum has non-trivial topological structure~[1--3]. Indeed, even
if set to zero by hand in the QCD Lagrange density, it will be
regenerated when contributions from instanton solutions are included
in the path integral. Its space-time integral does not necessarily
vanish but is proportional to the winding number (Pontryagin index) of
the field configuration. The $\theta$-term will then contribute
intrinsically non-perturbative CP violation, i.e.\ its effects will be
invisible to perturbation theory. Since no CP violation has been
observed in QCD, $\theta_{\rm eff}$ must be very small.

This can be quantified by examining the electric dipole moment of the
neutron, $d_n$: the CP violation induced by the $\theta$-term leads to
a neutron electric dipole moment of order \mbox{$|d_n| \approx
|\theta_{\rm eff}| 10^{-16} e \; {\rm cm}$}~\cite{d_ntheory}, which must be
compared to the current experimental limit \mbox{$d_n<0.63\times
10^{-25} e \; {\rm cm}$}~\cite{d_nexp}. Therefore $|\theta_{\rm eff}| \lesssim
10^{-9}$, naturally leading to the question: why is CP violation in
QCD so small? This is known as the ``strong CP problem''. \\

The axion provides a very natural solution to the strong CP
problem. It was realized that the $\theta$-term could be absorbed by
making a redefinition (an axial rotation) of the quark
fields~\cite{thooft}. If the quarks have zero mass the Lagrange
density will be unchanged except for the removal of the $\theta$-term,
and theories with differing values of $\theta_{\rm eff}$ all represent the same
physics. In essence, the $\theta$-term can be rotated away using the
global U(1) axial symmetry of the model. However, if the quarks have
non-zero mass then this rotation will introduce complex phases into
the quark mass matrix and the theory will still be CP-violating.

Peccei and Quinn~\cite{PQ} pointed out that if a new global axial
symmetry, a Peccei-Quinn (PQ) symmetry, is introduced then it could be
used to remove the $\theta$-term instead. When this PQ symmetry is
dynamically broken by the vacuum structure it will result in a
pseudo-Nambu-Goldstone boson known as the axion~\cite{axion}. It is
only a ``pseudo''-Nambu-Goldstone boson because the PQ symmetry is not
exact --- it is explicitly broken by the triangle anomaly providing a
non-perturbative axion-gluon coupling. This axion-gluon coupling has
two effects. Firstly, it will provide a non-zero axion mass due to
mixing with the pion, which is approximately given by
\be M_a = \frac{f_{\pi} m_{\pi}}{4 \langle \phi_a \rangle} 
\sqrt{\frac{4 m_u m_d}{(m_u+m_d)^2}} \left[1+ {\cal O}(m_{u,d}/m_s) \right] 
\approx 0.6 \times 10^{-3} {\rm eV} \left[ \frac{10^{10} {\rm GeV}}{f_a} 
\right], \label{eq:axionmass}\ee
where $m_u$, $m_d$ and $m_s$ and the up, down and strange quark masses
respectively, $f_{\pi}$ and $m_{\pi}$ are the pion decay constant and
the pion mass, and $\phi_a$ is the axion field.
Secondly, the axion-gluon coupling introduces an effective term into
the Lagrange density of the same form as the $\theta$-term,
Eq.(\ref{eq:thetaterm}), so that the CP-violating terms become
\be {\cal L}_{\theta\;{\rm eff}} = \left(\theta_{\rm eff} - \frac{\phi_a}{f_a}\right) 
\frac{\alpha_s}{8 \pi} F^{\mu \nu \, a} \tilde{F}_{\mu \nu}^a, \ee
where $f_a$ is the axion decay constant.  However, the potential for
$\phi_a$ is also a function of $\left(\theta_{\rm eff} -
\phi_a/f_a\right)$ and so the axion field relaxes to a
vacuum-expectation-value (VEV) given by $\langle \phi_a \rangle = f_a
\theta_{\rm eff}$. The $\theta$-term is canceled and the strong CP problem is
solved. \\

The experimental bounds on the existence of the axion are already
rather strict~\cite{pdb}. The non-observation of an axion in collider
experiments and rare decays (e.g.\ quarkonium decays) rules out models
where the PQ scale ($f_a$) is of the order of the electroweak
scale. However, these bounds can always be avoided by increasing the
PQ scale~\cite{ksvz,dfsz}, or equivalently reducing the axion mass,
thereby reducing the axion's couplings to known particles.

In order to constrain this ``invisible axion'' one must consider
astrophysical constraints~\cite{astrorev}. Since a low mass axion is
expected to be emitted during star cooling, $f_a$ may be constrained
by insisting that the axion does not significantly alter the observed
stellar evolution. Stars in globular clusters are the most sensitive
to these effects~[12-13].
Additionally, the neutrino signal from
SN 1987A indicates that it is cooled mainly by neutrino emission rather
than by emission of an ``invisible axion''~\cite{SN1987A}. Together
these observations place a limit of roughly $f_a \gtrsim 10^9 \; {\rm
GeV}$ (translating via Eq.(\ref{eq:axionmass}) to $M_a \lesssim 0.01
\; {\rm eV}$).

Intriguingly, at scales just above this limit the axion is seen to be
a good dark matter candidate. Indeed, it was shown in
Ref.\cite{pqscale_prediction} that in the standard thermal scenario,
and many inflationary models, the dark matter axion's PQ scale is
predicted to be $f_a \approx 3 \times 10^{10} \; {\rm GeV}$. If the PQ
scale becomes too much larger the axion contribution to dark matter
may become {\em too great}, thereby over-closing the universe and thus
providing an upper limit on $f_a$.  However, this upper bound is very
model dependent. We will see later that the main results of this letter
do not depend on the fine details of the axion mass limits, but only
that the PQ scale be very large. \\

In this letter, we will briefly describe the PQ symmetric
Next-to-Minimal Supersymmetric Standard Model (NMSSM), which is the
minimal supersymmetric extension of Standard Model that can provide an
axion. We will examine the Higgs boson mass spectrum of the model and
see that the lightest pseudoscalar Higgs boson is the ``invisible
axion'', and will subsequently be unobservable at colliders for the
foreseeable future. However, we will show that in order to keep the
mass-squared of the lightest scalar Higgs boson positive, one must
constrain the heavy Higgs bosons to lie in a very specific mass
window. We will provide one-loop expressions for this mass window in a
very good approximation. Therefore, this model provides a prediction
for the heavy Higgs boson masses which may be confirmed or ruled out at
the next generation of colliders. \\

\mytitle{The PQ Symmetric NMSSM}

One model that provides an axion is the PQ symmetric
NMSSM~[15--18]; this has the same field content as the
Minimal Supersymmetric Standard Model (MSSM) except for the inclusion
of an extra Higgs singlet superfield $\hat{S}$. Its superpotential is
given, in an obvious notation, by
\be 
W=\hat{u}^c \, \mathbf{h_u} \hat{Q} \hat{H}_u 
-\hat{d}^c \, \mathbf{h_d} \hat{Q} \hat{H}_d 
-\hat{e}^c \, \mathbf{h_e} \hat{L} \hat{H}_d 
+\lambda \hat{S}(\hat{H}_u \hat{H}_d). 
\label{eq:superpotential}
\ee
The usual Higgs--higgsino mass term $\mu \hat{H}_u \hat{H}_d$ seen in
the MSSM has been replaced by the term $\lambda \hat{S}(\hat{H}_u
\hat{H}_d)$ coupling the new singlet Higgs field, $\hat{S}$, to the Higgs
doublets, $H_d$ and $H_u$, where $\lambda$ is a dimensionless
parameter. The Higgs--higgsino mass term will be recovered when the
scalar component, $S$, of the new singlet superfield gains a 
VEV of $\langle S \rangle = \mu /\lambda$.

In the MSSM, the dimensionful parameter, $\mu$, is constrained to be
of the order of the electroweak scale in order to give the correct
pattern of electroweak symmetry breaking, even although it has no {\em
a priori} relation to the electroweak scale. The question of why two
seemingly unrelated scales should be the same is known as the
``$\mu$-problem''~\cite{mu-problem}. The original formulation of the NMSSM was
intended to answer this question by dynamically linking the scale
$\mu$ to a VEV of a Higgs field, $S$, and thereby to the electroweak
scale.

The superpotential, Eq.(\ref{eq:superpotential}), has no dimensionful
couplings and exhibits a U(1) PQ symmetry, which will be carried over
into the Lagrange density. In the MSSM this PQ symmetry is {\em
explicitly} broken by the Higgs-higgsino mass term $\mu \hat{H}_u
\hat{H}_d$; in the PQ symmetric NMSSM the PQ symmetry is only
{\em dynamically} broken when $S$ gains a non-zero VEV, giving rise to
a near massless pseudo--Nambu--Goldstone boson --- the
axion\footnote{The axion is only a ``pseudo''--Nambu--Goldstone boson
since the PQ symmetry is explicitly broken by the triangle anomaly,
giving it a small mass, Eq.(\ref{eq:axionmass}).}. Therefore the PQ
symmetric NMSSM is the minimal supersymmetric extension of the
Standard Model that can provide an axion. In fact, it is a
supersymmetric version of the DFSZ axion model~\cite{dfsz}.

The axion constraints mentioned in the introduction must also be
applied here and so models with $\langle S \rangle$ of the order of
the electroweak scale are ruled out. In the more usual formulation of
the NMSSM this is avoided by adding a term $\frac{1}{3} \kappa
\hat{S}^3$ to the superpotential; this {\em explicitly} breaks the PQ
symmetry, giving the `axion' a mass and avoiding the
constraints. Here, in order to preserve a near massless axion, we
insist that $\langle S \rangle \gtrsim 10^9 \; {\rm GeV}$. Therefore,
the PQ symmetric NMSSM no longer links $\langle S
\rangle$ to the electroweak scale and {\em cannot} be considered as a
solution to the $\mu$-problem. Since $\mu$ must remain of order the
electroweak scale, $\lambda=\mu/\langle S \rangle$ must be very small
and the $\mu$-problem is re-expressed as: why is $\lambda$ so small? 
We will not attempt to answer this question here. 

The axion within the context of the NMSSM has also been discussed in
Ref.\cite{nonpqaxion}. In that study, the term $\frac{1}{3} \kappa
\hat{S}^3$ was included in the superpotential, explicitly breaking the
PQ symmetry, but it was pointed out that in the limit where the soft
supersymmetry breaking parameters associated with $\lambda$ and
$\kappa$ vanish, the model will contain an additional approximate
$U(1)_R$ symmetry. This symmetry is dynamically broken by the
vacuum, giving rise to an `R-axion'. Unfortunately the mass of this
R-axion becomes rather large, forbidding its use in solving the strong
CP problem, but nevertheless the model has interesting
phenomenological consequences. \\

The superpotential, Eqn.(\ref{eq:superpotential}), leads to the
tree--level Higgs potential~\cite{ellis}:
\be V = V_F+V_D+V_{\rm soft}, \label{eq:Hpot} \ee
with
\ba
V_F&=& |\lambda S|^2 (|H_u|^2+|H_d|^2) + |\lambda H_uH_d|^2 , \label{eq:HpotF} \\
V_D &=& \frac{1}{8} \bar g^2 (|H_d|^2-|H_u|^2)^2
 +\frac{1}{2}g^2|H_u^{\dagger}H_d|^2 , \label{eq:HpotD} \\
V_{\rm soft}&=&m_{H_u}^2|H_u|^2 + m_{H_d}^2|H_d|^2 + m_S^2|S|^2
+ [\lambda A_{\lambda}SH_uH_d+\textrm{h.c.} ], 
\label{eq:HpotS}
\ea
where $\bar g = \sqrt{g^2+g^{\prime 2}}$ with $g$ and $g^{\prime}$
being the gauge couplings of $SU(2)_L$ and $U(1)$ interactions
respectively, and we have adopted the notation $H_uH_d \equiv
\epsilon_{\alpha \beta}(H_u)^{\alpha}(H_d)^{\beta}=H_u^+H_d^--H_u^0H_d^0$.  
The first two terms, $V_F$ and $V_D$, are the $F$ and $D$ terms
derived from the superpotential in the usual way, while $V_{\rm soft}$
contains the soft supersymmetry--breaking parameters $A_{\lambda}$,
$m_{H_u}$, $m_{H_d}$ and $m_S$.

The vacuum of the model may be rendered neutral by a suitable
application of a $SU(2)_L \times U(1)_Y$ gauge transformation, and 
rendered real by exploiting the PQ symmetry. The vacuum is then given
by
\be 
        \langle H_d \rangle = \frac{1}{\sqrt{2}} {v_d \choose 0}, 
\qquad  \langle H_u \rangle = \frac{1}{\sqrt{2}} {0 \choose v_u}, 
\qquad  \langle S \rangle = \frac{1}{\sqrt{2}} v_s, \label{eq:potmin} \ee
with $v_s$, $v_u$, and $v_d$ real and positive.  The requirement for
this vacuum to be a local minimum provides three relations, linking
the three soft mass parameters to the three VEVs of the Higgs fields:
\ba
m_{H_d}^2&=& \frac{1}{8} {\bar g}^2 (v_u^2-v_d^2) -\frac{1}{2} \lambda^2 v_u^2
 + \frac{1}{\sqrt{2}} A_{\lambda} \lambda v_s \frac{v_u}{v_d} 
- \frac{1}{2} \lambda^2 v_s^2, \label{eq:mind}\\
m_{H_u}^2&=& \frac{1}{8} {\bar g}^2 (v_d^2-v_u^2) -\frac{1}{2} \lambda^2 v_d^2 
+ \frac{1}{\sqrt{2}} A_{\lambda}\lambda v_s \frac{v_d}{v_u} 
- \frac{1}{2} \lambda^2 v_s^2, \label{eq:minu} \\
m_S^2 &=& -\frac{1}{2}\lambda^2v^2
+\frac{1}{\sqrt{2}} \lambda A_{\lambda} \frac{v_uv_d}{v_s}; \label{eq:mins}
\ea
as usual, we have written $v\equiv \sqrt{v_u^2+v_d^2}$. \\

The extra singlet fields mix with the Higgs doublet fields, increasing
the rank of the scalar and pseudoscalar mass--squared mixing matrices
by one each. After an initial rotation of the Higgs doublet fields by
an angle $\beta$, defined as usual via $\tan \beta \equiv v_u/v_d$ and
outlined in detail in Ref.\cite{me_nmssm}, the $2 \times 2$
pseudoscalar mass matrix is given by
\be M_A^2 \left(
\begin{array}{ccc}
1 & \frac{1}{2} \sin 2 \beta \cot \beta_s \\
\frac{1}{2} \sin 2 \beta \cot \beta_s & \frac{1}{4} \sin^2 2 \beta \cot^2 \beta_s 
\end{array}
\right). \ee
In analogy to $\tan \beta$, we have also defined $\tan \beta_s \equiv
v_s/v$; due to the requirement that \mbox{$\langle S \rangle \gtrsim 10^9 \;
{\rm GeV}$}, $\tan \beta_s$ will be very large, and therefore $\cot
\beta_s$ very small.  In the above, we have {\it defined} the 
upper--left entry of the pseudoscalar mass--squared mixing matrix to
be $M_A^2$. This new mass parameter replaces the soft
supersymmetry--breaking parameter $A_{\lambda}$ and becomes the mass
of the MSSM pseudoscalar Higgs boson as the MSSM limit is approached,
i.e.\ $\cot \beta_s \to 0$ with $\mu$ fixed. This treatment allows
higher order loop corrections to be absorbed directly into the
definition of $M_A$. Including one-loop top/stop corrections, it is
related to $A_{\lambda}$ by
\be
M_A^2 = \frac{2 \mu}{\sin 2 \beta} A_{\lambda}
-\frac{3 h_t^2}{16 \pi^2} A_t F(\msta^2,\mstb^2)
\ee
where $h_t=\sqrt{2}m_t/(v \sin \beta)$ is the top-quark Yukawa
coupling and $A_t$ is its associated soft supersymmetry-breaking mass
parameter. The function $F$ is given by
\be
F(\msta^2,\mstb^2)=\frac{1}{\msta^2-\mstb^2} 
\left[ \msta^2 \log \left( \msta^2/Q^2 \right) 
     - \mstb^2 \log \left( \mstb^2/Q^2 \right) \right] -1
\ee
and $m_t$, $\msta$, $\mstb$ are the top and stop masses, with $Q$ the
renormalization scale. \\

This pseudoscalar mass-squared matrix is easily diagonalized,
revealing two mass eigenstates, which will be denoted $A_1$ and $A_2$
with the label assigned in order of increasing mass. The first of
these, $A_1$, is the massless Nambu--Goldstone boson associated with
the dynamical breaking of the PQ symmetry --- the axion. The PQ
symmetry ensures that it will be massless even after the inclusion of
loop corrections; it only gains a very small mass via non-perturbative
mixing with the pion, as described earlier. The heavier mass
eigenstate, $A_2$, has mass
\be
M_{A_2}^2 = M_A^2 (1 + \frac{1}{4} \sin^2 2 \beta \cot^2 \beta_s).
\ee
Since $\cot \beta_s \lesssim 10^{-7} \; {\rm GeV}$ the heavy
pseudoscalar Higgs boson reproduces the mass of the MSSM pseudoscalar
with a deviation less than one part in $10^{14}$. \\

Similarly, the symmetric $3 \times 3$ scalar Higgs mass-squared matrix
is
\be
	M^2 = M_0^2 + \Delta,
\ee
where the entries of the tree-level contribution, $M_0^2$, can be written as
\ba
{[M_0^2\,]}_{11}&=& M_A^2 + (M_Z^2-\mu^2 \cot^2 \beta_s) \sin^2 2 \beta \\
{[M_0^2\,]}_{12}&=&- \frac{1}{2} (M_Z^2-\mu^2 \cot^2 \beta_s) \sin 4 \beta \\
{[M_0^2\,]}_{13}&=&- \frac{1}{4}M_A^2\sin4\beta \cot \beta_s \\
{[M_0^2\,]}_{22}&=&M_Z^2\cos^2 2 \beta + \mu^2 \cot^2 \beta_s \sin^2 2 \beta \\
{[M_0^2\,]}_{23}&=&\frac{1}{2}(4\mu^2-M_A^2\sin^22\beta)\cot \beta_s \\
{[M_0^2\,]}_{33}&=&\frac{1}{4}M_A^2\sin^22\beta \cot^2 \beta_s
\ea
$\Delta$ denotes higher order corrections to the scalar Higgs mass
matrix~\cite{vloop,roman}. Including one-loop top/stop corrections
these are given by~\cite{roman}
\ba
\hspace*{-0.7cm} {\Delta}_{11}&=& \frac{3h_t^2}{8\pi^2} m_t^2 \left[
s_{\beta}^2 \log \left[ \frac{\msta^2\mstb^2}{m_t^4} \right]
-8 \frac{a^4}{s_{\beta}^2}
K_1(\msta^2,\mstb^2) 
+8 a^2 
K_2(\msta^2,\mstb^2)
\right], \label{eq:nlo11} \\
\hspace*{-0.7cm} {\Delta}_{12}&=& \frac{3h_t^2}{8\pi^2} m_t^2 \left[
s_{\beta} c_{\beta} \log \left[ \frac{\msta^2\mstb^2}{m_t^4} \right]
-8 \frac{a^3 b}{s_{\beta}^2} K_1(\msta^2,\mstb^2) 
+4 \frac{a}{s_{\beta}} (a c_{\beta} + b s_{\beta})
K_2(\msta^2,\mstb^2)
\right], \\
\hspace*{-0.7cm} {\Delta}_{13}&=& \frac{3h_t^2}{16\pi^2} \left[
-\sqrt{2} \mu s_{\beta} c_{\beta}
F(\msta^2,\mstb^2)
+ 8 m_t^2 a^2 b \frac{c_{\beta}}{s_{\beta}^2} K_1(\msta^2,\mstb^2) 
\right. \nn \\ &&  \hspace*{6.5cm}  \left.
- 4 m_t^2 a \frac{c_{\beta}^2}{s_{\beta}} K_2(\msta^2,\mstb^2)
\right] \sqrt{2} \mu \cot \beta_s, 
\\
\hspace*{-0.7cm} {\Delta}_{22}&=& \frac{3h_t^2}{8\pi^2} m_t^2 \left[
c^2_{\beta} \log \left[ \frac{\msta^2\mstb^2}{m_t^4} \right]
-8 a^2 b^2 \frac{1}{s^2_{\beta}} K_1(\msta^2,\mstb^2) 
+8 a b \frac{c_{\beta}}{s_{\beta}} K_2(\msta^2,\mstb^2)
\right], \\
\hspace*{-0.7cm} {\Delta}_{23}&=& \frac{3h_t^2}{16\pi^2} \left[
\sqrt{2} \mu c^2_{\beta} F(\msta^2,\mstb^2)
+ 8 m_t^2 a^3 \frac{c_{\beta}}{s^2_{\beta}} K_1(\msta^2,\mstb^2) 
\right. \nn \\ && \hspace*{6.5cm} \left.
- 4 m_t^2 a c_{\beta} K_2(\msta^2,\mstb^2)
\right] \sqrt{2} \mu \cot \beta_s, \\
\hspace*{-0.7cm} {\Delta}_{33}&=& -\frac{3h_t^2}{2\pi^2} 
m_t^2 \mu^2 a^2 c^2_{\beta} \cot^2 \beta_s
K_1(\msta^2,\mstb^2), \label{eq:nlo33}
\ea
where $s_{\beta} \equiv \sin\beta$, $c_{\beta} \equiv \cos\beta$, 
$a=(-\mu c_{\beta}+A_t s_{\beta})/\sqrt{2}$, and 
$b=(\mu s_{\beta}+A_t c_{\beta})/\sqrt{2}$,
and the functions $K_1$ and $K_2$ are
\ba
K_1(\msta^2,\mstb^2)&\equiv&K(\msta^2,\mstb^2)/(\msta^2-\mstb^2)^2, \\
K_2(\msta^2,\mstb^2)&\equiv&(K(\msta^2,\mstb^2)+1)/(\msta^2+\mstb^2),
\ea
with
\be
K(\msta^2,\mstb^2)\equiv F(\msta^2,\mstb^2)
- \frac{1}{2} \log \left[ \frac{\msta^2\mstb^2}{m_t^4} \right].
\ee
Closed form expressions for the scalar Higgs boson mass eigenvalues
can be obtained by diagonalizing $M^2$. However, these results are
rather lengthy and unilluminating, and will not be reproduced
here. 

Fortunately, these exact expressions are not needed due to the very small
size of \mbox{$\cot\beta_s \sim {\cal O}(10^{-7})$}. Notice that the
mass-squared matrix takes the form
\be M^2=\left(
\begin{array}{ccc}
A_{11} &A_{12} & C_1 \cot\beta_s\\
A_{12} &A_{22} & C_2 \cot\beta_s\\
C_1 \cot\beta_s&C_2 \cot\beta_s& B \cot^2\beta_s
\end{array}
\right). \ee
This is true not only at tree-level but also when higher orders are
included.  We may reduce this matrix to block diagonal form by
applying a unitary transformation defined by the $3 \times 3$ matrix
\be V^{\dagger} = \left(
\begin{array}{cc}
1 \hspace{-0.16cm} 1 - \frac{1}{2} \cot^2\beta_s \Gamma \Gamma^{\dagger}
& - \cot\beta_s \Gamma \\
\cot\beta_s \Gamma^{\dagger}
& 1 - \frac{1}{2} \cot^2\beta_s \Gamma^{\dagger} \Gamma
\end{array}
\right)+{\cal O}(\cot^4\beta_s), \ee
with
\be \Gamma= 
\left(C_1 A_{22} - C_2 A_{12}, \, - C_1 A_{12} + C_2 A_{11} \right)/{\rm det} \; A.
\ee
Applying this transformation gives the simple form
\ba \lefteqn{VM^2V^{\dagger} =} \hspace{0.5cm} &  \nn \\ 
& \hspace*{-1cm} \left(
\begin{array}{ccc}
A_{11} + C_1 \Gamma_1 \cot^2\beta_s & 
\hspace{-0.7cm} A_{12} +\frac{1}{2} (C_1 \Gamma_2  + C_2 \Gamma_1 ) 
\cot^2\beta_s & \hspace{-1.4cm} 0 \\
A_{12} +\frac{1}{2} (C_1 \Gamma_2  + C_2 \Gamma_1 ) \cot^2\beta_s & 
\hspace{-0.7cm} A_{22} + C_2 \Gamma_2 \cot^2\beta_s & \hspace{-1.4cm} 0 \\
0      & \hspace{-0.7cm} 0      & \hspace{-1.4cm} (B - C_2 \Gamma_2 - C_1 
\Gamma_1) \cot^2\beta_s
\end{array}
\right) \label{eq:bdm2} \\ &\hspace{12.5cm} +{\cal O}(\cot^3\beta_s). \nn \ea

The upper-left block consists of the usual MSSM scalar Higgs boson
mass-squared matrix (to any desired number of loops) plus corrections
of order $\cot^2\beta_s$. Consequently, the two heavier states, $H_2$
and $H_3$, are rather uninteresting; the MSSM scalar Higgs masses,
like a heavy pseudoscalar, will be recovered with corrections of only
one part in $10^{14}$, which is neither experimentally observable, nor
theoretically reliable since unincluded higher order corrections will
present much larger deviations. This was to be expected since our
NMSSM parameter choice is approaching the MSSM limit. \\

\mytitle{A prediction for $M_A$}

The lightest Higgs boson, whose mass-squared is given by the
lower-right entry, is rather more interesting. Its mass is suppressed
by $\cot\beta_s$, making it effectively massless at current collider
energies, but its couplings to known particles, which mainly arise
from the mixing with the other scalar Higgs bosons\footnote{The
Lagrangian of the model also contains new direct couplings of the new
singlet state to known particles but these are also suppressed by at
least one order of $\cot\beta_s$.}, will also be tiny. Subsequently,
this state would be unobservable at high energy colliders for the
foreseeable future, and the low energy phenomenology would appear
indistinguishable from the MSSM.

However, the expression for the lightest scalar mass shows interesting
structure. Inserting the tree-level values into the lower-right entry
of Eq.(\ref{eq:bdm2}) gives the tree level mass-squared
\be
M_{H_1}^2 = 
 \mu^2 \tan^2 2\beta \cot^2\beta_s (
-(x^2+y^2)(x^2-1)^2
+y^2 \cos^2 2\beta
)/(xy)^2
\label{eq:mh12tree} \ee
where $x \equiv M_A \sin2\beta/(2 \mu)$ and $y \equiv M_Z
\sin2\beta/(2 \mu)$. 

\begin{figure}[h]
\begin{center}
\includegraphics[scale=0.4]{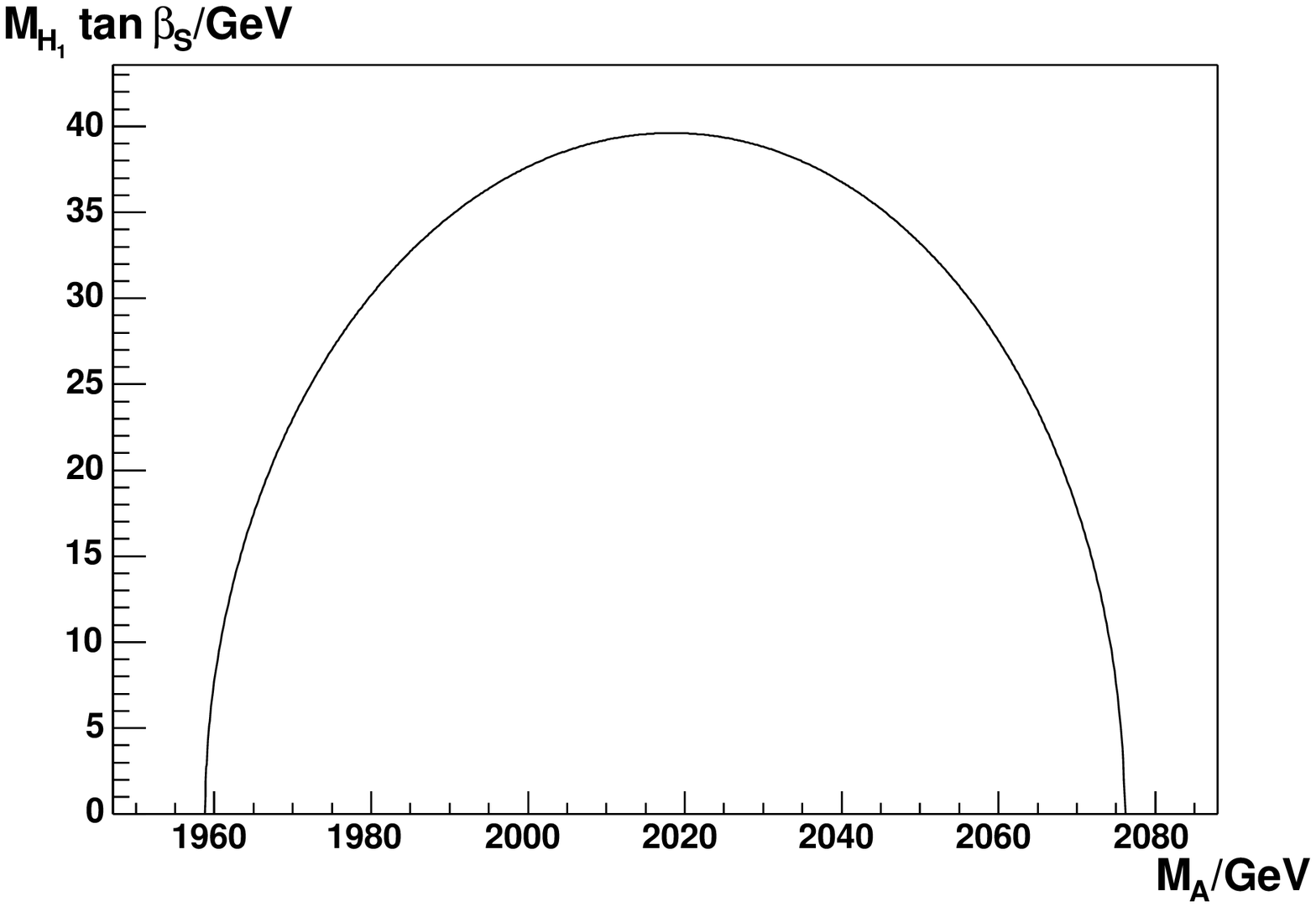}
\includegraphics[scale=0.4]{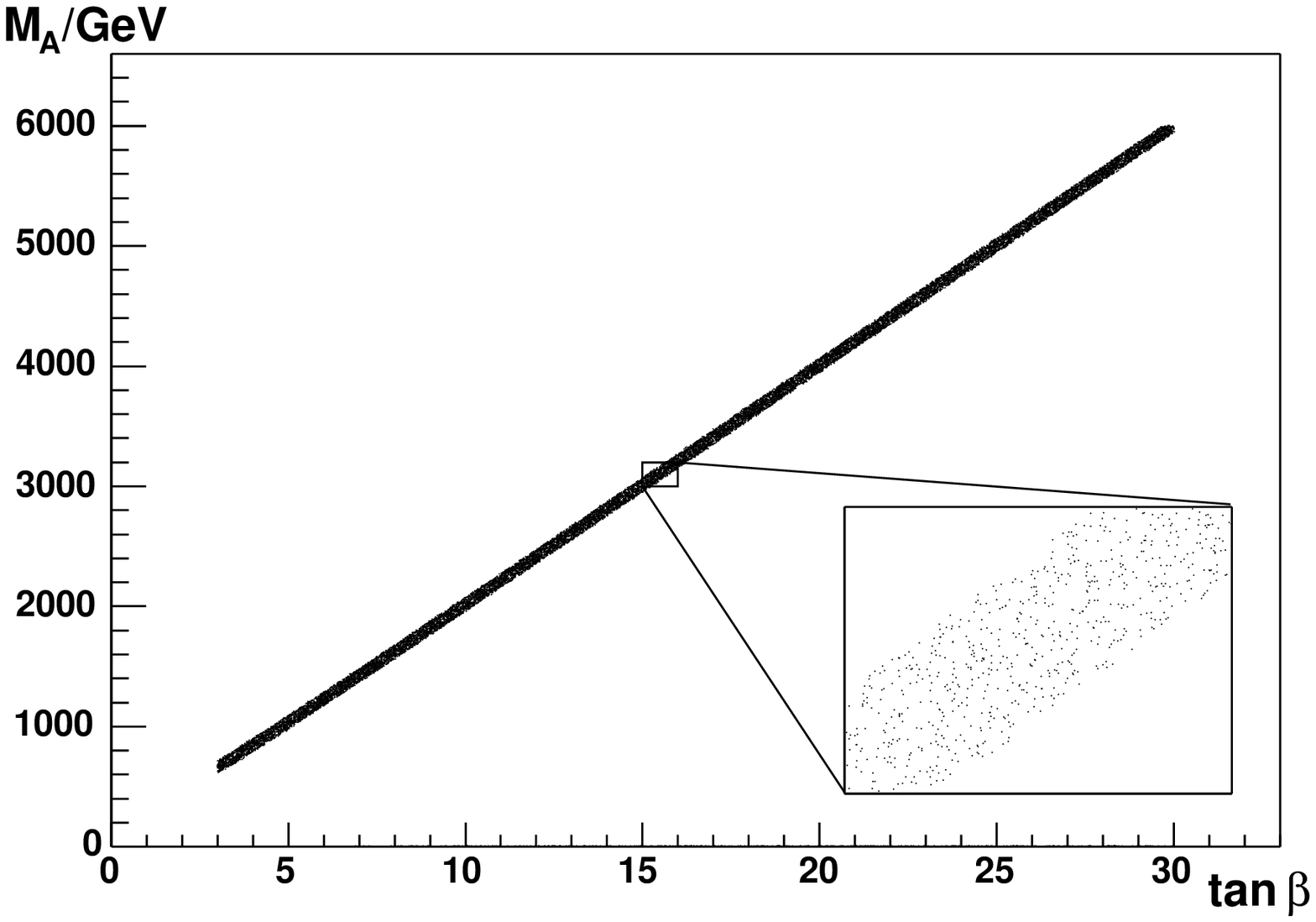}
\caption{\it Left: 
The dependence of the lightest scalar Higgs mass (normalized by $\tan
\beta_s$) on $M_A$, for $\tan \beta = 10$ and $\mu=200 \; GeV$. 
Beyond the points where the curve meets the axis the mass-squared
becomes negative and the vacuum unstable.  Right: The distribution of
scenarios with physically acceptable vacua, with $M_A$ chosen randomly
between $0$ and $6$~TeV and $\tan\beta$ chosen randomly between $3$
and $30$. The vacuum structure constrains the value of $M_A$ to lie
close to approximately $\mu \tan \beta$. The blow-up allows individual
scenario points to be seen.}
\label{fig:matbscan}
\end{center}
\end{figure}

This mass-squared must be positive in order to have a physically
acceptable theory. If it is negative, the Higgs potential will be
unbounded from below and the vacuum unstable. However, only the last
term in the brackets of Eq.(\ref{eq:mh12tree}) is positive;
$M_{H_1}^2$ will become negative for both high and low values of
$M_A$, and a stable vacuum will be achieved only for a small range
around $x \approx 1$. This is also true when loop corrections are
included, as shown in Fig.(\ref{fig:matbscan},~left).

To demonstrate this we examined $10^6$ different scenarios, with $M_A$
and $\tan \beta$ chosen randomly between $0$ to $6$~TeV and $3$ to
$30$ respectively. we calculated the one-loop mass spectrum and, for
every scenario with a stable vacuum, plotted a single point on the
$M_A$--$\tan\beta$ plane of Fig.(\ref{fig:matbscan}, right). We
discarded scenarios with unstable vacua.
It is immediately evident that the physically acceptable scenarios all
lie within a small band around $M_A \approx 2 \mu / \sin 2\beta
\approx \mu \tan \beta$. 

Therefore the PQ symmetric NMSSM with a large expectation value of the
new singlet field makes a {\it prediction} for the masses of the heavy
Higgs bosons. This prediction is potentially falsifiable, or
verifiable, at the next generation of colliders. Furthermore, as long
as $\cot \beta_s$ is small, the positivity or negativity of
$M_{H_1}^2$ is independent of $\cot \beta_s$, and consequently the
prediction of the heavy Higgs boson masses is also independent of the
value of $\cot \beta_s$. Therefore, if after measuring $\mu$ and
$\tan\beta$ at a future collider, the heavy pseudoscalar mass is not
found to lie close to $\mu \tan \beta$ then this model is ruled out
for {\em all large values of the singlet expectation value.} 
Alternatively, if the mass prediction were found to hold, it would
provide very tantalizing, albeit indirect, evidence for the PQ
symmetric NMSSM as a solution to the strong CP problem and for the PQ
axion as a source of dark matter. \\

In order to compare the values of $\mu$ and $\tan \beta$ with $M_A$ at
the next generation of colliders, the vacuum stability bounds must be
made more precise. Since $M_{H_1}^2=0$ with $M_{H_1}^2$ given by
Eq.(\ref{eq:mh12tree}) is only a cubic in $x^2$, it can be solved to
give closed form analytical expressions for the tree-level
boundary. Throwing the third non-physical solution away, we find
\be
x^2_{\rm max/min}=1 - \frac{1}{3} (1+y^2)(1-\cos \gamma \pm \sqrt{3} \sin \gamma) 
+\Delta_{\pm}, \label{mabound}
\ee
where
\be
\gamma = \frac{1}{3} \tan^{-1} \left( -\frac{\sqrt{4(1 + y^2)^6 - (2 (1 + y^2)^3 - 27y^2 \cos^22\beta)^2}}{2 (1 + y^2)^3 - 27y^2 \cos^22\beta} \right),
\ee
and $\Delta_{\pm}$ represents the higher order corrections.

Since the one-loop top/stop contributions to $\Delta_{\pm}$,
Eqs.(\ref{eq:nlo11}-\ref{eq:nlo33}), are independent of $M_A$,
\mbox{$M_{H_1}^2=0$} remains cubic in $M_A^2$ when these corrections are 
included and we can still find a closed form solution for the
limits. However, these expressions are long and complicated, and once
again such complexity is not needed here. Instead we expand the
one-loop corrections as a series in the small parameter $y$ and
discard terms of ${\cal O}(y^3)$. This gives
\ba
\Delta_{\pm}&=&
\frac{1}{64 \mu^2} 
\left( 
\mp \frac{8}{y} \frac{s_{2\beta}^2}{c_{2\beta}} \Delta_{22}
+ [32 \Delta_{23} 
+ 8 s_{2\beta}^2 \Delta_{22} 
- 16 s_{2\beta} c_{2\beta} \Delta_{12} ]
\right. \nn \\ && \left.
\qquad\qquad \mp y\frac{1}{s_{2\beta}^2 c_{2\beta}} 
\left[
  8 s_{2\beta}^2 c_{2\beta}^4 \Delta_{11}
- 8 s_{2\beta}^3 (1+s_{2\beta}^2) c_{2\beta} \Delta_{12}
+   s_{2\beta}^4 (3+s_{2\beta}^2) \Delta_{22}
\right.\right. \nn \\ && \left.\left.
\qquad\qquad\qquad\qquad\qquad 
-32 s_{2\beta} c_{2\beta}^3 \Delta_{13}
+16 s_{2\beta}^2 c_{2\beta}^2 \Delta_{23}
+32 c_{2\beta}^2 \Delta_{33}
\right]
\right. \nn  \\ && \qquad\qquad
+16y^2 \left[
    s_{2\beta}^2 c_{2\beta}^2 \Delta_{11}
+   s_{2\beta}^3 c_{2\beta} \Delta_{12}
- 2 s_{2\beta} c_{2\beta} \Delta_{13}
\right]
\Bigg) + {\cal O}(y^3),
\ea
where $\Delta_{ij}$ are given by Eqs.(\ref{eq:nlo11}--\ref{eq:nlo33}).

This approximation is rather good. The non-observation of
supersymmetry to date requires that $\tan \beta \gtrsim 3$ and $\mu
\gtrsim 80$~GeV, giving $y \lesssim 0.34$. The discarded terms will
therefore alter the one-loop corrections by at most a few percent. For
more typical MSSM parameter choices, $y$ will be even smaller; e.g.\
for the Snowmass reference point SPS~1a~\cite{snowmass}, $\tan
\beta=10$ and $\mu \approx 350$~GeV, giving $y \approx 0.026$

A large $\tan \beta$ expansion of the tree-level result gives a very
approximate, but rather useful, ``rule of thumb'':
\be M_A \approx \mu \tan \beta \pm M_Z. \ee \\

The coupling of the lightest scalar Higgs boson to electrons may also
be restricted by astrophysical data, allowing more stringent limits to
be placed on the PQ scale. Just as for the axion, $H_1$ will be
produced during the cooling of globular--cluster stars if its mass is
below about $10$~keV. The maximum value of the $H_1$ mass seen in
Fig.(\ref{fig:matbscan},~left) is realized\footnote{More accurately,
making a series expansion in the small parameter $y$, the maximum
(tree-level) value of $M_{H_1}$ is found at
\mbox{$x=1+\frac{1}{2} y^2 \cos^2 2\beta +{\cal O}(y^4)$.}} at $x \approx 1$;
inserting this into Eq.(\ref{eq:mh12tree}) gives
\be M_{H_1}^{\rm max} \approx \mu \sin 2\beta \cot \beta_s, \ee
so the limits from star cooling cannot be avoided if
$\langle S \rangle \gtrsim 2 \mu \sin 2\beta \times 10^7 \gtrsim
10^{10}$~GeV, where for the last inequality we have made the
reasonable assumption that $\mu \lesssim 1$~TeV and $\tan \beta >3$.

Above this scale one must respect the limits on the coupling of the
lightest scalar Higgs boson to electrons~\cite{glob-scalar},
\be g_{H_1e} \lesssim 1.3 \times 10^{-14}. \ee
In the NMSSM it is easy to see that $g_{H_1e} \approx m_e/\langle S
\rangle$ and so this translates into a lower bound on the PQ
scale. Combining this with the requirement that the $H_1$
mass be less than $10$~keV for this lower bound to apply, {\em
excludes} the values
\be 
2 \mu \sin 2\beta \times 10^7 
\lesssim \langle S \rangle 
\lesssim 4 \times 10^{10} {\rm GeV.}
\ee
Allowing the maximum and minimum values of $\mu$ and $\tan \beta$
respectively, only a rather small range of $\langle S \rangle$ values
is unequivocally ruled out. However, as $\mu$ and $\tan \beta$ are
allowed to move toward less extreme values, the excluded range becomes
larger and soon overlaps with that disallowed by emission of the
axion from globular--cluster stars, i.e.\ $\langle S \rangle \gtrsim
10^9$~GeV. \\

Finally, since the model is supersymmetric, the extra neutral singlet
superfield also contains a higgsino, which will be manifest as an
extra neutralino --- the lightest supersymmetric particle (LSP) of the
model. Once again, the large value of the PQ scale leads to it having
a very small mass and being almost totally decoupled from the other
particles. To a good approximation, its mass is given by $M_{\rm LSP}
\approx \mu \cot^2 \beta_s$, which, for $\mu \approx 10^3$~GeV and
$\langle S \rangle \approx 10^{11}$~GeV translates to\footnote{It is
intriguing to note that this mass lies not too far from the expected
neutrino mass scale.} $M_{\rm LSP} \approx 3 \times 10^{-6}$~eV. In
contrast to the scalar and pseudoscalar Higgs bosons, R-parity
conservation prevents the LSP being emitted during star cooling, so it
provides no further astrophysical limits. \\

\mytitle{Summary \& Conclusions}

In this letter, we have discussed the Next-to-Minimal Supersymmetric
Standard Model (NMSSM) with an explicit Peccei-Quinn (PQ)
symmetry. This model is the minimal supersymmetric extension of the
Standard Model that can provide an axion. This axion is a
pseudo--Nambu--Goldstone boson associated with the dynamical breaking
of the PQ symmetry, and is manifest in this model as the lightest
pseudoscalar Higgs boson; it can be used to solve the strong CP
problem of QCD and is a dark matter candidate. The stellar evolution
of globular cluster stars and the neutrino signal from SN 1987A
provide a lower bound on the PQ breaking scale $\gtrsim 10^9 \; {\rm GeV}$.

We have shown that in this limit simple expressions for the NMSSM
Higgs boson masses can be obtained. The heavy and intermediate mass
Higgs bosons have masses and couplings indistinguishable from those of
the corresponding MSSM. The lightest scalar and pseudoscalar (the
axion) decouple from the other particles and will be invisible to
future collider searches.

However, we have demonstrated that in order that the theory have a
stable vacuum, i.e.\ in this case that the lightest scalar mass-squared
be positive, the heavy mass scale $M_A$ must lie within approximately
$\mu \tan \beta \pm M_Z$. We have presented analytic expressions for
these limits on $M_A$ to one-loop top/stop accuracy.

If, at a future collider, $M_A$ were found to be outside this range,
then the PQ symmetric NMSSM would be ruled out {\em for all values of
the PQ scale}. This is not an unreasonable event; the restriction on
$M_A$ is unlikely to occur by chance without some other organizing
principle. For example, {\em all} of the Snowmass MSSM reference
points~\cite{snowmass}, which are considered a representative sample
of MSSM scenarios, fail this criterion.  It is important to stress
that only the axion associated with this particular model would be
ruled out; an axion could still be present via some other mechanism,
and axion search experiments, such as CAST~\cite{cast}, the U.S.\
Axion Search (Livermore)~\cite{livermore} and the Kyoto search
experiment CARRACK~\cite{carrack} would still be very important.

On the other hand, if the heavy Higgs boson mass scale were seen to
obey the bound given by Eq.(\ref{mabound}) we would have very exciting
circumstantial evidence for the existence of an NMSSM axion. Then the
role of the axion search experiments would become even more crucial.\\

\mytitle{Acknowledgments}

The authors would like to thank P.M.~Zerwas for his continual support
and encouragement. DJM is grateful to W.~Buchm\"uller for helpful
discusions, and M.~Pl\"umacher for a critical reading of the
manuscript. RN is grateful to V.~Rubakov and H.B.~Nielsen for useful
remarks and comments. \\

\mytitle{References} 
\end{document}